# A Modified FRW Metric to Explain the Cosmological Constant

Serkan Zorba

Department of Physics and Astronomy

Whittier College

13406 Philadelphia Street

Whittier, CA 90608

szorba@whittier.edu

Abstract:

One of the most outstanding problems of the standard model of cosmology today is the problem of cosmological constant/dark energy. It corresponds to about 73 per cent of the energy content of the universe gone missing. I hereby postulate a modified FRW metric for our universe, which animates a universe spinning rigidly but very slowly with an angular frequency that is equal to the Hubble constant. It is shown by a simple argument that in such a universe there will be an overlooked rotational energy whose average value is identically equal to the matter-energy content of this universe as observed by a coordinate observer.

## 1. INTRODUCTION

One of the most significant unsolved problems of modern cosmology is the problem of the cosmological constant (Weinberg 1989, Carroll 2001). We still don't know its origin, the reason for its extreme smallness, and the reason for its conflict with the quantum field theoretic calculations, all of which perplex cosmologists to this day (Abbott 1998, Garriga & Vilenkin 2000, Steinhardt & Turok 2006). Various models and ideas have been proposed to explain the origin of the cosmological constant, including dark energy, quintessence, the anthropic principle, and phantom energy (Caldwell, Dave & Steinhardt 1998, Penrose 2007, Beck & Mackey 2008, Caldwell, Kamionkowski & Weinberg 2003). However, none of these models has been able to settle the problem in a satisfactory way (Carroll 2001, Beck & Mackey 2008).

The prime candidate for the cosmological constant is dark energy. Dark energy is a mysterious negative-pressure energy, giving rise to a repulsive force. It is believed to be the reason behind the observed acceleration of the universe. The repulsive force associated with dark energy increases linearly with distance. Dark energy is usually associated with the vacuum energy of the universe (Penrose 2007). However, quantum field theoretic calculation of the total vacuum energy - with an upper cutoff frequency imposed by the Planck scale - gives a value of about 120 orders of magnitude higher than the observed cosmological constant (Lyre 2003, Sahni & Starobinsky 2000). This is indeed a grand discrepancy that has baffled physicists greatly (Hawking 2001, Hartle 2003, Cheng 2005).

In this paper, I will advance a simple and less-mysterious model of the universe to resolve the problem of the cosmological constant/dark energy: a rotationally dynamic universe, which offers a natural and simple explanation for the accelerated expansion of the universe and the origin of Hubble's law.

It must be noted right at the outset that there is no uncontroversial observational evidence, yet, that our universe is rotating. In fact, in today's standard cosmology, the universe is assumed to be non-rotational. The CMBA data gleaned so far suggest more or less a homogeneous and isotropic universe, albeit with some problematic irregularities to explain. Nevertheless, there is, of course, the scandalous cosmological constant/dark energy problem with strong observational evidence. The latter amply warrants investigation of different, if not seemingly outrageous, universe models.

The motivation behind my rotational model to resolve the cosmological constant/dark energy problem is threefold: (i) The positive repulsive force - which varies linearly with distance - conjectured for the dark energy; (ii) the form of Hubble law, in which the speed of receding galaxies varies linearly with distance and the Hubble constant playing the role of an angular frequency; and (iii) the observed so-called cosmic "axis of evil" and non-Gaussian properties of the Cosmic Microwave Background Anisotropies (CMBA). In my view, all of the foregoing are evocative of a rotating universe, and justify studying models of a rotational universe, which are largely simple modifications of the hitherto successful standard FRW model.

A globally rotating universe is consistent with general relativity (GR). Rotating universe models have been studied before starting with Gödel (1949). Most rotational models utilize Bianchi cosmological models but not all (e.g., see Su & Chu 2009).

## 2. PROPOSED MODEL AND RESULTS

I postulate the following modified Friedmann-Robertson-Walker (FRW) line element, where a slow forced-vortex type rotation with a constant angular frequency equal to the Hubble constant ($\omega = H_0 = 2.28 \times 10^{-18}\ rad\ s^{-1}$) is incorporated into the classic FRW metric. Since the postulated rotation is so slow, the directional scale factors (as encountered in Bianchi cosmological models) in the metric I propose are chosen to be equal to each other (namely, $R_1 \cong R_2 \cong R_3 = R$). The latter is justified by the CMBA observations. Thus, my metric represents a homogeneous, quasi-isotropic, shear-free, and extremely slowly rotating universe. So it still keeps the spirit and gist of the standard FRW metric. The latter is also the reason why I will work in the spherical coordinates.

$$d\tau^2 = \left(1 - \frac{\omega^2 R^2 r^2 sin^2\theta}{c^2}\right)dt^2 - R^2\left[\frac{dr^2}{c^2(1-kr^2)} + \frac{r^2}{c^2}d\theta^2 + \frac{r^2 sin^2\theta}{c^2}d\phi^2\right] + \frac{2\omega R^2 r^2 sin^2\theta}{c^2}d\phi dt. \quad (1)$$

Here $R = R(t)$ is the effective scale factor of the universe and changes very slowly with time, k is the curvature constant, and $\omega$ is the angular frequency of the rotation. Since the speed with which a galaxy at a distance $r$ from the center of rotation will be moving is given by $v = \omega r$, by the constancy of the speed of light $c$, we would have $\omega = c/R_0$, where $R_0$ is the Hubble radius of the universe (Kragh 1999). The Hubble radius represents our universe's gravitational horizon as explained by Melia and coworkers in a recent paper (2010).

Let us derive the Friedmann differential equation for the proposed metric using the full machinery of GR (Mould 1994, Hobson, Efstathiou & Lasenby 2006).

The calculation of the corresponding metrics, Christoffel symbols, Riemann curvature and Ricci tensors can be found under Appendix A.

The six surviving components of the Ricci tensor are found to be

$$R_{00} = \frac{3\ddot{R}}{R} - (\ddot{R}R + 2\dot{R}^2 + 2kc^2)\frac{\omega^2 r^2 sin^2\theta}{c^2} \; ;$$

$$R_{11} = -(\ddot{R}R + 2\dot{R}^2 + 2kc^2)\frac{1}{c^2(1-kr^2)} \; ;$$

$$R_{22} = -(\ddot{R}R + 2\dot{R}^2 + 2kc^2)\frac{r^2}{c^2} \; ;$$

$$R_{33} = -(\ddot{R}R + 2\dot{R}^2 + 2kc^2)\frac{r^2 sin^2\theta}{c^2} \; ;$$

$$R_{03} = R_{30} = (\ddot{R}R + 2\dot{R}^2 + 2kc^2)\frac{\omega r^2 sin^2\theta}{c^2}.$$

The covariant form of the energy-momentum tensor for an ideal fluid, which is used in the FRW model of the standard model of cosmology, is given by

$$T_{\mu\nu} = \left(\rho + \frac{p}{c^2}\right)u_\mu u_\nu - p g_{\mu\nu}, \qquad (2)$$

where $\rho$ is the average matter-energy density, p is the average pressure, u's are the four-velocity vectors of the galaxy ensemble.

For our model of a spinning universe, the energy component $p_0$ of the covariant momentum four-vector will be conserved as an object moves over a geodesic path because our metric has a very weak time dependence, and hence can be found as follows:

$p_\mu = g_{\mu\nu} p^\nu$, where $p^\nu = E(u^\nu) = (E, 0, 0, 0)$. Note here that in comoving coordinates, the contravariant four velocity is given by $u^\nu = (1, u^1, u^2, u^3) = (1, 0, 0, 0)$, and $E = E_{inertial}$ is the matter-energy content with respect to an inertial observer and is given in terms of the rest mass energy, $E_0$, as

$$E = E_0\left(1 - \frac{\omega^2 R^2 r^2 \sin^2\theta}{c^2}\right)^{-1/2} = E_0/\sqrt{1-x^2},$$

where $x^2 = \frac{\omega^2 R^2 r^2 \sin^2\theta}{c^2}$ is used for short hand. The maximum value $x$ (the reduced distance) can take is 1, which corresponds to the Hubble radius. (Note that at the present time $R^2 r^2 \sin^2\theta = R_0^2$, where $R_0$ is the Hubble radius of the universe, and $c = \omega R_0$.)

Using our covariant metric components, we obtain the covariant energy as observed by the coordinate observer at the axis of rotation

$$E_{coord} = p_0 = g_{\mu\nu}p^\nu = g_{00}p^0 + g_{03}p^3 = g_{00}p^0 = E(1-x^2) = E_0\sqrt{1-x^2}.$$

Therefore the energy expression due to rotation is the difference between the matter-energy content observed by the inertial observer and that of the coordinate observer.

$$E_{rot} = E_{inertial} - E_{coord} = E_0\left((1-x^2)^{-1/2} - \sqrt{1-x^2}\right) = \frac{E_0 x}{\sqrt{\frac{1}{x^2} - 1}}.$$

This is the energy expression whose average value the coordinate observer will overlook when he/she surveys the rotating universe. The average values of these energy functions on the interval [0, 1] are found as follows (see Fig. 2 below)

$$\bar{E}_{inertial} = \int_0^1 E_0/\sqrt{1-x^2}\, dx = E_0 \pi/2.$$

$$\bar{E}_{coord} = \int_0^1 E_0\sqrt{1-x^2}\, dx = E_0 \pi/4.$$

Thus $\bar{E}_{inertial} = 2\bar{E}_{coord}$. In other words, $\bar{E}_{rot} = \bar{E}_{coord}$. Therefore, in a rotating universe, the coordinate observer will overlook an average matter energy that corresponds to the average energy of the rotation of the universe. Curiously, this difference is identically equal to the average matter-energy content of the coordinate observer's universe. One must take that extra energy into account much in the same way the dwellers of the Earth take into consideration the inertial forces (centrifugal and Coriolis effects) to properly explain their observations of, for instance, large scale weather phenomena.

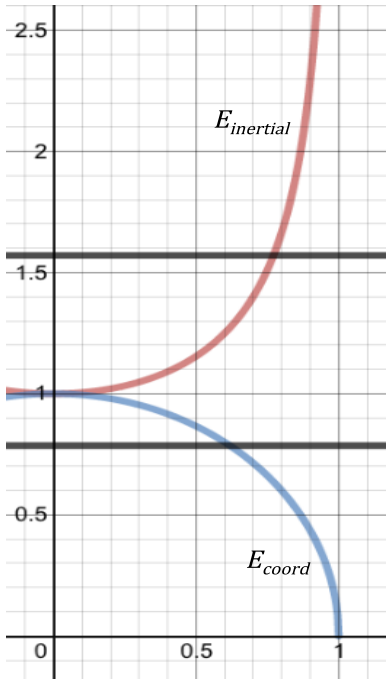

FIG. 1. Matter-energy content of a rotating universe (expressed as a ratio over the rest-mass energy) is plotted as a function of the reduced distance $x$ as observed by an inertial observer (top curve), and a coordinate observer (bottom curve). The horizontal black lines represent the average values of these respective functions. We note that the average energy observed by the inertial observer is exactly twice that observed by the coordinate observer. The difference is due to rotation. The energy overlooked by the coordinate observer is identically equal to the observed energy content of his/her universe.

Thus the relevant average total energy on the average expressed in terms of matter-energy density in a rotating system must be $\rho_{tot} = 2\rho$.

The correct energy-momentum tensor with the rotational energy taken into account will be

$$T_{\mu\nu} = \{2\rho + \tfrac{p}{c^2}\}u_\mu u_\nu - p g_{\mu\nu}. \tag{3}$$

For short hand, let us call $\rho_r = 2\rho$.

One can now determine the right hand side of the Einstein field equation $T_{\mu\nu} - \tfrac{1}{2}T g_{\mu\nu}$ of our slowly rotating universe. The trace of the energy-momentum tensor is found to be $T = \rho_r c^2 - 3p$. The six nonzero components are found to be (writing $x$ explicitly, to match the Ricci tensor expressions given earlier)

$$T_{00} - \tfrac{1}{2}T g_{00} = \tfrac{c^2}{2}(\rho_r c^2 + 3p) + \tfrac{(\rho_r c^2 - p)}{2}\omega^2 R^2 r^2 \sin^2\theta \ ;$$

$$T_{11} - \tfrac{1}{2}T g_{11} = \tfrac{(\rho_r c^2 - p)}{2}\tfrac{R^2}{(1-kr^2)} \ ;$$

$$T_{22} - \tfrac{1}{2}T g_{22} = \tfrac{(\rho_r c^2 - p)}{2} R^2 r^2 \ ;$$

$$T_{33} - \tfrac{1}{2}T g_{33} = \tfrac{(\rho_r c^2 - p)}{2} R^2 r^2 \sin^2\theta \ ;$$

$$T_{30} - \tfrac{1}{2}T g_{30} = T_{03} - \tfrac{1}{2}T g_{03} = -\tfrac{(\rho_r c^2 - p)}{2}\omega R^2 r^2 \sin^2\theta \ ;$$

Here we note that when $\omega = 0$, the Ricci tensor and the energy-momentum tensor both reduce to those of FRW, as expected.

Now plugging the Ricci tensor and the energy-momentum tensor for a spinning FRW universe into the Einstein field equation without any cosmological term, $R_{\mu\nu} = \kappa(T_{\mu\nu} - \tfrac{1}{2}T g_{\mu\nu})$, where $\kappa = -\tfrac{8\pi G}{c^4}$, we get two independent equations:

$$\tfrac{3\ddot{R}}{R} - (\ddot{R}R + 2\dot{R}^2 + 2kc^2)\tfrac{\omega^2 r^2 \sin^2\theta}{c^2} = \tfrac{\kappa c^2(\rho_r c^2 + 3p)}{2} + \tfrac{\kappa c^2(\rho_r c^2 - p)}{2}\omega^2 R^2 r^2 \sin^2\theta. \tag{4}$$

$$(\ddot{R}R + 2\dot{R}^2 + 2kc^2) = -\frac{\kappa R^2 c^2 (\rho_r c^2 - p)}{2}. \tag{5}$$

Using Eq. 4 and Eq. 5 we arrive at the Friedmann equation for our spinning universe:

$$\frac{\dot{R}^2}{R^2} = \frac{8\pi G}{3}\rho_r - \frac{kc^2}{R^2} = \frac{8\pi G}{3}(2\rho) - \frac{kc^2}{R^2} = \frac{8\pi G}{3}\rho + \frac{8\pi G}{3}\rho - \frac{kc^2}{R^2}. \tag{6}$$

As seen, we obtained an extra term (as compared with the non-spinning FRW model) naturally without assuming any cosmological constant. Let us scrutinize that extra term using a symbolic expression for the current value of the observed density of the universe. The current cosmic density is approximately equal to the critical density of the universe. Thus, $\rho \cong \rho_c = \frac{3H_0^2}{8\pi G}$, where $H_0$ is the Hubble constant at the present epoch (Suyu et al. 2010, Tegmark et al. 2004). Furthermore, since $\omega = c/R_0$, our extra term becomes

$$\frac{8\pi G}{3}\rho = \frac{8\pi G}{3}\left(\frac{3H_0^2}{8\pi G}\right) = H_0^2.$$

In other words, our spinning FRW-universe Friedmann equation can be now written as

$$\frac{\dot{R}^2}{R^2} = \frac{8\pi G}{3}\rho + H_0^2 - \frac{kc^2}{R^2}. \tag{7}$$

The standard Friedmann equation for a non-spinning FRW universe with an artificially added cosmological constant term is given by (Hobson et al. 2006)

$$\frac{\dot{R}^2}{R^2} = \frac{8\pi G}{3}\rho + \frac{1}{3}\Lambda c^2 - \frac{kc^2}{R^2}, \tag{8}$$

where $\Lambda$ is the cosmological constant.

Comparing Eq. 8 (non-spinning FRW) with Eq. 7 (spinning FRW), we can see that $\frac{1}{3}\Lambda c^2 = H_0^2$, giving us

$$\Lambda = \frac{3 H_0^2}{c^2}. \tag{9}$$

Now, the standard (non-spinning) FRW model with the artificially added cosmological constant also has $\Lambda = \frac{3H_0^2}{c^2}$. We therefore see that the herein proposed spinning-universe model reproduces exactly the result obtained from the FRW metric as used in the Einstein equation with an artificially added cosmological constant.

## 3. QUASI-NEWTONIAN DERIVATION

In many a textbook on GR, it is a fairly common approach to derive the Friedmann differential equation for the standard FRW model following a quasi-Newtonian argument within an overall general relativistic framework (Cheng 2005, Rindler 2006, Rohlf 1994). Let us apply this approach to our rotating universe model to see if we will obtain the cosmological constant term as a result of the rotational energy of the universe.

Consider a spherical volume of radius $r$, where $r$ is much larger than the average inter-galaxy distance, and much smaller than the size of the universe. As the universe is observed to be expanding, we can write

$r = r_0 a(t)$,

where $r_0$ is the size of our sphere at some arbitrary time, and $a(t)$ is a time-dependent scale factor. The Hubble's parameter, $H(t)$, is given by

$H(t) = \frac{\dot{a}}{a}$

Now, consider a galaxy of mass m that lies on the surface of the sphere. Assuming a pressureless matter, the conservation of energy requires (without the rotational energy being taken into account)

$$E_k + V = E_{tot}, \qquad (10)$$

where $E_{tot}$ is written in terms of the time-independent curvature parameter, $k$, of the universe as

$$E_{tot} \equiv -\frac{kc^2 m r_0^2}{2R_0^2}, \qquad (11)$$

where $R_0$ is the radius of the universe at the current epoch (Rohlf 1994). Since I postulate the existence of a slow rotation of the universe, Eq. 10 must be modified to be

$$E_k + V + \phi_{rot} = E_{tot}, \qquad (12)$$

where $\phi_{rot}$ is the energy contribution due to rotation. Eq. 12 can be written explicitly as

$$\frac{1}{2} m r_0^2 (\dot{a})^2 - \frac{4\pi m G r_0^2 a^2 \rho}{3} - \frac{1}{2} m (r_0 a)^2 \omega^2 = -\frac{kc^2 m r_0^2}{2R_0^2}, \qquad (13)$$

where $\omega$ is the angular frequency of the rotation - and according to my spinning universe model it is equal to the Hubble's constant $H_0$ at the present time - $G$ is the gravitational constant, $\rho$ is the average mass-energy density of matter in the universe, and $c$ is the speed of light in vacuum. See Appendix B for how a negative rotational energy, $\phi_{rot}$, is obtained. Rearranging Eq. 13, we get the Friedmann equation for my model

$$(\dot{a})^2 = \frac{8\pi G \rho a^2}{3} - \frac{kc^2}{R_0^2} + a^2 \omega^2. \qquad (14)$$

Written in terms of the Hubble's parameter, Eq. 14 becomes

$$H^2 = \frac{8\pi G \rho}{3} - \frac{kc^2}{R_0^2 a^2} + \omega^2. \tag{15}$$

Since the standard Friedmann equation, with the cosmological constant, is (Hobson et al. 2006)

$$H^2 = \frac{8\pi G \rho}{3} - \frac{kc^2}{R_0^2 a^2} + \frac{8\pi G}{3} \rho_\Lambda, \tag{16}$$

where $\rho_\Lambda$ is the so-called vacuum density, which is given in terms of the cosmological constant, $\Lambda$, as $\rho_\Lambda = \frac{c^2}{8\pi G} \Lambda$. Comparing Eq. 15 with Eq. 16, we find $\frac{8\pi G}{3} \rho_\Lambda = +\omega^2$, or in terms of the cosmological constant $\Lambda \frac{c^2}{3} = \omega^2$. Thus $\Lambda = \frac{3\omega^2}{c^2}$.

Since my model views the present value of the Hubble's constant as the angular frequency of the spinning universe,

$$\Lambda = \frac{3H_0^2}{c^2}, \tag{17}$$

which is the exact equation obtained from (Eq. 9) my full general relativistic derivation above. This is very significant because, like the general relativistic treatment, the quasi-Newtonian treatment of the rotating universe also naturally produces the cosmological constant term in the Friedmann equation.

## 4. FURTHER DISCUSSION

In a slowly rotating frame of reference, the galaxies will feel a centrifugal force given by $F_c = m\omega^2 r = mH_0^2 r$, which is the repulsive force originating from the assumed rotation of the universe, and is directly proportional with distance $r$.

A de Sitter universe, an empty universe model ($\rho = k = p = 0$), which is the working model for many inflationary cosmologies, has only the cosmological constant term to be dominant. Our Eq. 7 in a de Sitter model produces the scale factor expression as $R(t) \propto e^{\omega t} = e^{H_0 t}$ (Mould 1994).

A similar exponential "expansion" also results classically in a situation where the centrifugal force - with which we are associating the effect of a cosmological constant, $\Lambda$, in a de Sitter space - is the only dominant perceived force, as in a centrifuge. In the frame of reference of the object feeling only the centrifugal force, Newton's second law can be written as $\ddot{r} - \omega^2 r = 0$, which gives rise to a non-trivial solution of $r = r_0 e^{+\omega t}$, with $\omega = H_0 \propto \sqrt{\Lambda}$. The classical time-dependent scale factor here can be seen to be $e^{H_0 t}$.

It is clear that a de Sitter model with a cosmological constant term is nothing but a holding-onto the rotational energy of the universe even while stripping it out of its material content. In other words, according to my model, the cosmological constant corresponds to the rotational energy of the universe.

In addition to providing a natural and viable solution to the many seemingly mysterious inferences and observations of our universe, as demonstrated in this letter, a spinning universe model also presents another route for its verification and/or falsification. A spinning universe around a cosmic axis of rotation implies an overall alignment of the axes of rotation of galaxies by conservation of angular momentum and general relativistic precession effects (Rindler 2006). The homogeneity employed in the metric however implies that the direction of the axis is primary not its "position."

The analysis of WMAP data already reveals a preferred axis—the so-called cosmic "axis of evil"—with which the axes of rotation of most galaxies seem to line up (Schwarz et al. 2004, Land & Magueijo 2005, Longo 2007, 2011). Polarization vector measurements of quasars also seem to indicate a mysterious alignment of their rotation axes (Hutsemékers et al. 2008).

Furthermore, my model places an observational constraint of about $10^{-10}$ rad yr$^{-1}$ on the rotation of our universe. This compares well with a recent estimate (Su & Chu 2009).

As mentioned at the outset, rotating universe models were offered before. Gödel's metric is one of the most famous ones. Gödel did not produce the cosmological constant term from his approach. In fact, following the convention, he included the cosmological constant term artificially in his solution. However, Gödel's model also implied a rotation axis and - not having any observational support for a potential cosmic axis - he tried to justify the lack of order in the alignment of the then-observed galaxies by presciently pointing out that there might "exist various circumstances which would tend to blur the original order, or make it appear blurred" (Gödel 1949).

In summary, I proposed a modified FRW metric, which assumes a slowly spinning universe (with its angular frequency being equal to the Hubble's constant). The solution of the Einstein equation produces the cosmological constant term exactly. I posited that our universe appears to possess a mysterious dark energy because it is rotating, and the rotational energy of the universe is perceived by us as dark energy. If the universe is indeed rotating, then the standard model of cosmology would have to be modified. The logical antecedent of a spinning universe is a "Big Spin," with a colossal initial angular momentum, instead of a giant spatial expansion called the Big Bang. Spinning expands the universe with repulsive (centrifugal) forces, and naturally results in Hubble's law.

My model thus explains the following in a simple and natural (non-mysterious) way:

1- The origin of the cosmological constant, and hence the resolution of the problem of dark energy.

2- Hubble's law.

3- Accelerated expansion of the universe.

4- The repulsive nature of the force of the so-called dark energy, as well as its linear dependence on distance.

5- The cosmic "axis of evil," along which the rotation axes of galaxies and quasars seem to be aligned.

Although there is no conclusive observational evidence (yet) that our universe is rotating, the universe must be rotating precisely because a rotational universe does naturally and simply explain the foregoing five points, which are all grounded in observations and measurements.

## 5. CONCLUSIONS

I proposed a spinning universe model, which naturally explains many seemingly mysterious and difficult cosmological issues such as the origin and extreme small value of the cosmological constant/dark energy, its gravitational repulsive force, Hubble's law, and the so-called cosmic "axis of evil."

According to my model, the universe is rigidly rotating with an angular frequency equal to the Hubble's constant, and with a centrifugal force that is linearly proportional to distance. It implies that dark energy is not the "vacuum energy," but rather the rotational energy of the universe. My model thus has significant implications for the cosmological principle and the standard model of cosmology.

**Appendix A**

The metric components corresponding to the proposed line element (Eq. 1) are as follows.

$g_{00} = 1 - \frac{\omega^2 R^2 r^2 \sin^2\theta}{c^2}$ ; $g_{11} = -\frac{R^2}{c^2(1-kr^2)}$ ; $g_{22} = -\frac{R^2 r^2}{c^2}$ ; $g_{33} = -\frac{R^2 r^2 \sin^2\theta}{c^2}$ ;

$g_{03} = g_{30} = \frac{\omega R^2 r^2 \sin^2\theta}{c^2}$.

$g^{00} = 1$ ; $g^{11} = -\frac{c^2(1-kr^2)}{R^2}$ ; $g^{22} = -\frac{c^2}{R^2 r^2}$ ; $g^{33} = \omega^2 - \frac{c^2}{R^2 r^2 \sin^2\theta}$ ;

$g^{03} = g^{30} = \omega$.

The Christoffel symbols are calculated using

$\Gamma^{\sigma}_{\mu\nu} = \frac{1}{2} g^{\sigma\tau} \left( g_{\mu\tau,\nu} + g_{\tau\nu,\mu} - g_{\mu\nu,\tau} \right)$

The nonzero Christoffel symbols are calculated to be

$\Gamma^0_{00} = \frac{R\dot{R}\omega^2 r^2 \sin^2\theta}{c^2}$ ; $\Gamma^0_{03} = \Gamma^0_{30} = -\frac{R\dot{R}\omega r^2 \sin^2\theta}{c^2}$ ; $\Gamma^0_{11} = \frac{R\dot{R}}{c^2(1-kr^2)}$ ;

$\Gamma^0_{22} = \frac{R\dot{R}r^2}{c^2}$ ; $\Gamma^0_{33} = \frac{R\dot{R}r^2 \sin^2\theta}{c^2}$ ;

$\Gamma^1_{00} = (kr^2-1)\omega^2 r \sin^2\theta$ ; $\Gamma^1_{01} = \Gamma^1_{10} = \frac{\dot{R}}{R}$ ;

$\Gamma^1_{03} = \Gamma^1_{30} = (1-kr^2)\omega r \sin^2\theta$ ;

$\Gamma^1_{11} = \frac{kr}{(1-kr^2)}$ ; $\Gamma^1_{22} = (kr^2-1)r$ ; $\Gamma^1_{33} = (kr^2-1)r\sin^2\theta$ ;

$\Gamma^2_{00} = -\frac{\omega^2}{2}\sin 2\theta$ ; $\Gamma^2_{02} = \Gamma^2_{20} = \frac{\dot{R}}{R}$ ; $\Gamma^2_{03} = \Gamma^2_{30} = \frac{\omega}{2}\sin 2\theta$ ;

$\Gamma^2_{12} = \Gamma^2_{21} = \frac{1}{r}$ ; $\Gamma^2_{33} = -\frac{\sin 2\theta}{2}$ ;

$\Gamma^3_{00} = -\frac{2\dot{R}\omega}{R} + \frac{R\dot{R}\omega^3 r^2 \sin^2\theta}{c^2}$ ; $\Gamma^3_{01} = \Gamma^3_{10} = -\frac{\omega}{r}$ ; $\Gamma^3_{02} = \Gamma^3_{20} = -\omega\cot\theta$ ;

$\Gamma^3_{03} = \Gamma^3_{30} = \frac{\dot{R}}{R} - \frac{R\dot{R}\omega^2 r^2 \sin^2\theta}{c^2}$ ; $\Gamma^3_{11} = \frac{R\dot{R}\omega}{c^2(1-kr^2)}$ ; $\Gamma^3_{13} = \Gamma^3_{31} = \frac{1}{r}$ ;

$$\Gamma^3_{22} = \frac{RR\omega r^2}{c^2} \; ; \; \Gamma^3_{23} = \Gamma^3_{32} = \cot\theta \; ; \; \Gamma^3_{33} = \frac{RR\omega r^2 \sin^2\theta}{c^2} \; ;$$

The Riemann curvature tensor components are found using

$$R^\sigma_{\alpha\beta\delta} = -\Gamma^\sigma_{\alpha\beta,\delta} + \Gamma^\sigma_{\alpha\delta,\beta} - \Gamma^\tau_{\alpha\beta}\Gamma^\sigma_{\delta\tau} + \Gamma^\tau_{\alpha\delta}\Gamma^\sigma_{\beta\tau}.$$

And the Ricci tensor is found using

$$R_{\alpha\beta} = R^\delta_{\alpha\beta\delta}.$$

**Appendix B**

Quasi-Newtonian Derivation

Consider a sphere of mass $M$ and radius $r$, which is rotating around some axis with an extremely small angular frequency $\omega$. Now imagine a galaxy of mass $m$ at the surface of that sphere. In the reference frame of the galaxy, Newton's 2$^{nd}$ law (the equation of motion) will be given by the following expression

$$m\ddot{r} = -\frac{GmM}{r^2} + m\omega^2 r,$$

where the first term on the RHS is gravitational attraction and the second term is the fictitious (very real for the galaxy) outward-pulling centrifugal force. Integrating this equation produces

$$m\frac{\dot{r}^2}{2} = +\frac{GmM}{r} + \frac{m\omega^2 r^2}{2} + C,$$

where $C$ is the integration constant, which corresponds to the total energy content of the sphere. In other words, $C$ is nothing but $E_{tot}$ given in Eq. 12. So Eq. 12 can be written as

$$E_{tot} = E_k + V + \phi_{rot} = m\frac{\dot{r}^2}{2} - \frac{GmM}{r} - \frac{m\omega^2 r^2}{2}.$$

As can be seen the $\phi_{rot}$ is given by $-\frac{1}{2}m\omega^2 r^2 = -\frac{1}{2}m(r_0 a)^2\omega^2$.

This makes sense because the galaxy observer will feel not only an inward-pushing gravitational attraction, but also an outward-pulling tension-like centrifugal force due to rotation $\boldsymbol{F_c} = m\omega^2 \boldsymbol{r}$. Since $\nabla \times \boldsymbol{F_c} = 0$, it is a conservative force. Thus the galaxy observer will associate with it a potential (dark) energy of the form $\phi_{rot} = -\int F_c \cdot dr = -\frac{m\omega^2 r^2}{2}$. The latter can also be looked upon as a type of negative kinetic energy.

Comparing $\phi_{rot}$ with the gravitational potential energy (see Fig. 2 below), we see that the former creates a force to pull an object to infinity, while the gravitational potential energy creates a force to push the object to the origin. They thus are in a tug of war, in which the rotational potential energy will dominate, as the distance gets larger. Since $\omega = H_0$ in my model universe, $\phi_{rot}$ can also be expressed as

$$\phi_{rot} = -\frac{1}{2}m(r_0 a)^2 \omega^2 = -\frac{1}{2}mr_0^2 a^2 \left(\frac{\dot{a}}{a}\right)^2 = -\frac{1}{2}mr_0^2 \dot{a}^2.$$

Thus $\phi_{rot} \sim -\frac{1}{2}\dot{a}^2$. This is similar to the negative kinetic term as employed in phantom field models.

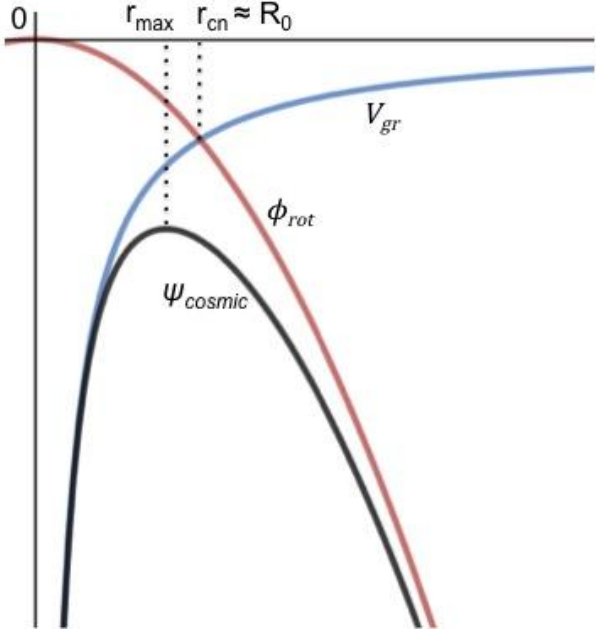

FIG. 2. Rotational potential, $\phi_{rot}$, and gravitational potential, $V_{gr}$, energies of the rotary model are co-plotted versus distance. Note how the strengths of the two energies coincide in almost today's era, represented by the Hubble radius $R_0$. The resulting cosmic potential $\psi_{cosmic}$ is also shown. We see that the universe rolled to the potential maximum (zero net-force) recently, beyond which universal acceleration starts. Today we are more or less at the cosmic potential hilltop. From there on our cosmic future will be all "downhill."